\documentclass[twocolumn,prl,showpacs,amsmath,amssymb]{revtex4}
\usepackage{graphicx}
\input epsf

\newcommand{\be}{\begin{equation}}
\newcommand{\ee}{\end{equation}}
\newcommand{\bea}{\begin{eqnarray}}
\newcommand{\eea}{\end{eqnarray}}
\newcommand{\nn}{\nonumber}

\newcommand{\p}{\partial}

\begin{document}

\preprint{}

\title{Topological Structures in Yang Mills Magneto-Fluids}
\author{Bindu A. Bambah}
\affiliation{School of Physics, \\University of Hyderabad, \\
Hyderabad,  Andhra Pradesh,  500 046, India }
\author{Swadesh M. Mahajan}
\affiliation{Institute of Fusion Studies,\\ University of Texas,
\\ Austin, Texas, 78712 U.S.A \\ and }
\author{Chandrasekher Mukku}
\affiliation{International Institute of Information Technology,\\
Hyderabad, Andhra Pradesh,  500 032\\ India}


\begin{abstract}
Knotted  configurations  supported by a Yang Mills fluid-field  system are suggested as a model for glueballs.
\end{abstract}
\maketitle
\section{Introduction}

The experimental discovery  that quark-gluon plasmas (QGP)  display features peculiar to
strongly coupled fluids \cite{expt} has generated tremendous interest in building
the  non-Abelian equivalents of the standard, routinely used Abelian fluid models like magneto hydrodynamics (MHD) \cite{bambah,Heinz,Van,Holm,manuel,Choquet-Bruhat,Kolb,shuryak,blaizot,jackiw1,jackiw2}.

In one such recent model\cite{bambah}, the dynamics of  a  hot relativistic quark gluon fluid (with a non-Abelian charge) was described  in terms of a generalized Yang Mills tensor  born out of the unification of the gauge field and the flow-field tensor. The fact that it is possible to define a single non-Abelian unified field for the quark-gluon system strongly suggests that one should explore the system  for topological structures endowed with properties such as linkages or knottedness of the fluid field lines
that are preserved under ideal dynamics. If found,
such a nonlinear stable soliton-like state could be identified with the wonted qcd object - the glueball
\cite{Glueball,Buniy,Nair}. The primary objective of this work is to construct  and elucidate precisely such states as solutions to the fluid model \cite{bambah}

In Abelian electrodynamics, the helicity of a vector field  is the standard
measure of the extent to which field lines coil around each other \cite{Moffat,Woltjer}.
The concept of helicity is trivially generalized  to non-Abelian fields, and can serve as an index of topological complexity for the knotted solutions we are seeking. It is no wonder that the notion of  helicity plays an important role in the study of plasma stabilty, and  has been shown to have an intimate connection with knot theory.

The fluid-field model presented in \cite{bambah} can be  loosely considered as a
non-Abelian generalization of the relativistic ( in directed as well as  in thermal energy)
fluid description of electromagnetic plasmas \cite{mahajan}. The fluid equations are derived from a perfect fluid energy momentum tensor : $T^{\mu\nu}=p\eta^{\mu\nu}+h U^{\mu}U^{\nu}$ with $p$ as the pressure, and the enthalpy density  $h=m n_{R} f (T)$  where $m$ and $n_{R}$ are, respectively, the rest frame density and inertial mass of the particles comprising the fluid. The statistical attributes of the fluid are represented by the temperature dependent factor $f(T)=f$. Interestingly enough, for the important class of homentropic fluids, $f$ appears in the equations of motion only as a multiplier to the fluid four velocity $U_{\nu}^a$ (changing $U_{\nu}^a$ to $fU_{\nu}^a$); evidently the velocity displayed here carries a non-Abelian index  ($a$).  Notice that the perfect fluid form for $T^{\mu\nu}$ holds for both quark and gluon fluids, that is, for the gluon fields that have acquired a temperature dependent mass - the temperature dependence goes to define the appropriate $f$ and the constant of proportionality may be viewed as the equivalent of the inertial mass in the expression for the enthalpy density. Naturally the massless gluon field is represented by the field  tensor $F^{\mu\nu}$

In Ref. \cite{bambah}, it was shown that  the Lorentz force equation for a non-Abelian fluid takes the form
\begin{equation}U^{\alpha}{}_{a}(\frac{m}{g}S^{a}{}_{\alpha\beta}+F^{a}{}_{\alpha\beta})=0. \label{nab}\end{equation}
In the equation of motion, the non-Abelian fluid tensor  $S^{a}{}_{\mu\nu}$
\begin{equation}
S^{\mu\nu}_{ a}= {\cal{D}}^{\mu}(fU^{\nu}_
{a})-{\cal{D}}^{\nu}(fU^{\mu}_{ a})-im f^2[U^{\mu}_{ b},U^{\nu}_{
c}],
\end{equation}where, ${\cal{D}}_{\mu}$ is the generalized non-Abelian covariant derivative,
${\cal{D}}_{\mu}=\partial_{\mu}-ig[A_{\mu}, ]-im[fU_{\mu}, ]$ appears on an equal footing  with the standard field tensor $F^{a}{}_{\alpha\beta}$. We are, thus,
 lead naturally to a unified "minimally" coupled potential for
hot non-Abelian fluids \be
Q^{\mu}_{a}=A^{\mu}_a+\frac{m}{g}fU^{\mu}, \ee  that generates its
 own unified  fluid- field gauge tensor
 \be
 M_{a}^{\mu\nu}=\p_{\mu} Q_a^{\nu}-\p_{\nu}
 Q^{\mu}_a+gc^{bc}_{a}Q^{\mu}_bQ^{\nu}_c.
 \ee
It is pertinent to realize that  $S^{a}{}_{\mu\nu}$ contains the  non-linear  flow-field coupling
through ${\cal{D}}_{\mu}$  that depends on  the Yang-Mills  connection $A^{a}{}_{\mu}$.

We now have the machinery to explicitly construct topological
fluid field solutions. Unlike the pure fluid  or the pure
Yang-Mills systems, the coupled system  will sustain solutions ( similar to Magnetohydrodynamics) in which the
fluid  carries the Yang-Mills field with it, i.e the Yang Mills field is frozen in with
the flow.
Although $SU(3)$ is the relevant group for the QGP, we solve here
for illustration, the simpler problem for the symmetry group SU(2).

We are interested in finding  topologically nontrivial, and spatially localized
solutions. The non-Abelian magneto fluid equation of motion(1)
suggests  $ \bf{M}_{\mu\nu}=0 $ to be a possible solution. In keeping
with assumed  localization of the solution, let us assume an interior and an
exterior region. The exterior region extends out to infinity and
applying traditional boundary conditions on fields at infinity, the proposed solution
requires $\bf{Q}_{\mu}\longrightarrow 0$ at spatial infinity.
Thus we can take $\bf{Q}_{\mu}=0$
(physical meaning will be dealt with later) in the entire exterior region.

The boundary between the interior
and exterior regions is (without loss of generality), a three sphere
and forms the overlap region for the interior and exterior
solutions. Since $\bf{Q}_{\mu}$ is a gauge connection, the interior
solution $\tilde{\bf{Q}}_{\mu}$ is related to the exterior solution
$\bf{Q}_{\mu}$ through a gauge transformation \be
\tilde{\bf{Q}}_{\mu}={\bf{\Omega}}
{\bf{Q}}_{\mu}{\bf{\Omega}}^{\dag}-\frac{i}{g}{\bf{\Omega}}\partial_{\mu}{\bf{\Omega}}^{\dag}.\label{vec}\ee It
is not difficult to see that since
${\bf{M}}_{\mu\nu}=g {\bf{F}}_{\mu\nu}+m {\bf{S}}_{\mu\nu}$ (being the
curvature of the generalized connection
${\bf{Q}}_{\mu}= {\bf{A}}_{\mu}+\frac{m}{g}f{\bf{U}}_{\mu}$, while $\bf{A}_{\mu}$ is
the Yang-Mills connection), transforms covariantly. The generalized
connection, by virtue of being a connection, transforms
inhomogeneously and implies that $f{\bf{U}}_{\mu}$, the velocity
vectors must transform covariantly. The inhomogeneous terms in
the transformation of ${\bf{Q}}_{\mu}$ are to be clubbed with the
transformation of the Yang-Mills connection ${\bf{A}}_{\mu}$.

Thus,  for the solution we are developing  ${\bf{Q}}_{\mu}=0$ in the exterior, and $\tilde{\bf{Q}}_{\mu}$ is
pure gauge in the interior (both imply ${\bf{M}}_{\mu\nu}=0$) The overlap region being a three sphere $\$^{3}$
then tells us that the group element ${\bf\Omega}$, belongs to the
homotopy type given by the maps ${\bf\Omega}:\$^{3} \longrightarrow
SU(2)$ having chosen the gauge group to be $SU(2)$. The group
manifold of $SU(2)$ is isomorphic to the three sphere and we are
left with the maps ${\bf\Omega}:\$^{3}\longrightarrow \$^{3}.$ Such maps
are labelled by an integer, the "winding number" (n) of the
topological solution. Therefore our goal is to find an ${\bf\Omega}$ such
that its winding number is nonzero with the implication that the exterior
solution $\bf{Q}_{\mu}=0$ cannot be extended into the interior.

For a pure gauge field, the winding (or, Pontryagin) number is
simply given by \be n=\frac{1}{24\pi^{2}}\int d^{3}x
\epsilon^{ijk}Tr[({\bf\Omega}\partial_{i}{\bf\Omega}^{\dag})({\bf\Omega}\partial_{j}{\bf\Omega}^{\dag})({\bf\Omega}\partial_{k}{\bf\Omega}^{\dag})].\label{n}\ee
Since the interior solution is simply given by \be
\tilde{\bf{Q}}_{\mu}=\frac{-i}{g}{\bf\Omega}\partial_{\mu}{\bf\Omega}^{\dag},\label{sol}\ee
and ${\bf\Omega}:\$^{3}\longrightarrow\$^{3}$, we automatically find
that the solution satisfies $\tilde{\bf{Q}}_{0}=0$. While in the
exterior, we have required $\bf{Q}_{\mu}=0$. A discussion of these
conditions will be given below.

To construct an explicit "pure gauge" solution
whose winding number is nonzero, we will borrow from the study of
instantons \cite{jackiwrebbi} in pure Yang-Mills theories. Taking
\be
{\bf\Omega}(x)=\frac{|{\vec{x}}|^{2}-1}{1+|{\vec{x}}|^{2}}+\frac{2i{\vec{\sigma}\cdot\vec{x}}}{1+|{\vec{x}}|^{2}}\label{gauge}\ee
where $\vec\sigma$ are the Pauli matrices, it is easy to see that the
$SU(2)$ gauge components of $\bf{Q}_{\mu}$ are given by

 \be
{\bf{Q}}^1=\frac{-4}{g(\vec{x}^2+1)^2}((\frac{1}{2}(1-|\vec{x}|^2)+x^2)\hat{\vec{x}}+(xy+z)\hat{\vec{y}}+(xz-y)\hat{\vec{z}})
\ee \be
{\bf{Q}}^2=\frac{-4}{g(1+\vec{x}^2)^2}((xy-z)\hat{\vec{x}}+(\frac{1}{2}(1-|\vec{x}|^2)+y^2)\hat{\vec{y}}+(yz+x)\hat{\vec{z}})
\ee \be
{\bf{Q}}^3=\frac{-4}{g(1+\vec{x}^2)^2}((xz+y)\hat{\vec{x}}+(yz-x)\hat{\vec{y}}+(\frac{1}{2}(1-|\vec{x}|^2)+z^2)\hat{\vec{z}})
\ee
 As we have seen above, the time component of ${\bf{Q}}$,
$Q^{a}{}_{0}=0$. It is easy to see that $ \int_{\$^3} {\bf
Q^1}\wedge{\bf Q^2}\wedge {\bf Q^3}=
\int\frac{64}{g^3(1+\vec{x}^2)^3}dx\wedge dy\wedge
dz=\frac{2\pi^2}{g^3}$. From equation \ref{sol}, the fact that the
SU(2) one form is  $\vec{Q}=\sigma_iQ^i$ and the properties of the
product of three  $\sigma$ matrices,we can see that $\int_{\$^3}
{\bf Q^1}\wedge {\bf Q^2}\wedge {\bf Q^3}=\frac{2\pi^2}{g^3}n$,
where n is given in eqn.\ref{n}. Thus
 the winding number of this fluid field knot is $n=1$. We
illustrate the nature of this solution by plotting in figs.1,2 and 3
(with a composite plot in fig.4), the surfaces (in toroidal
coordinates) on which the $Q^i$ lie.

\begin{figure}[htbp] \epsfxsize=4cm \epsfbox{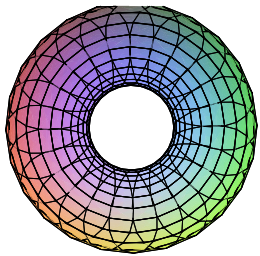}
\caption{shows the surface on which \\ $Q_1$ lies.}
\end{figure}
\begin{figure}[htbp] \epsfxsize=4cm \epsfbox{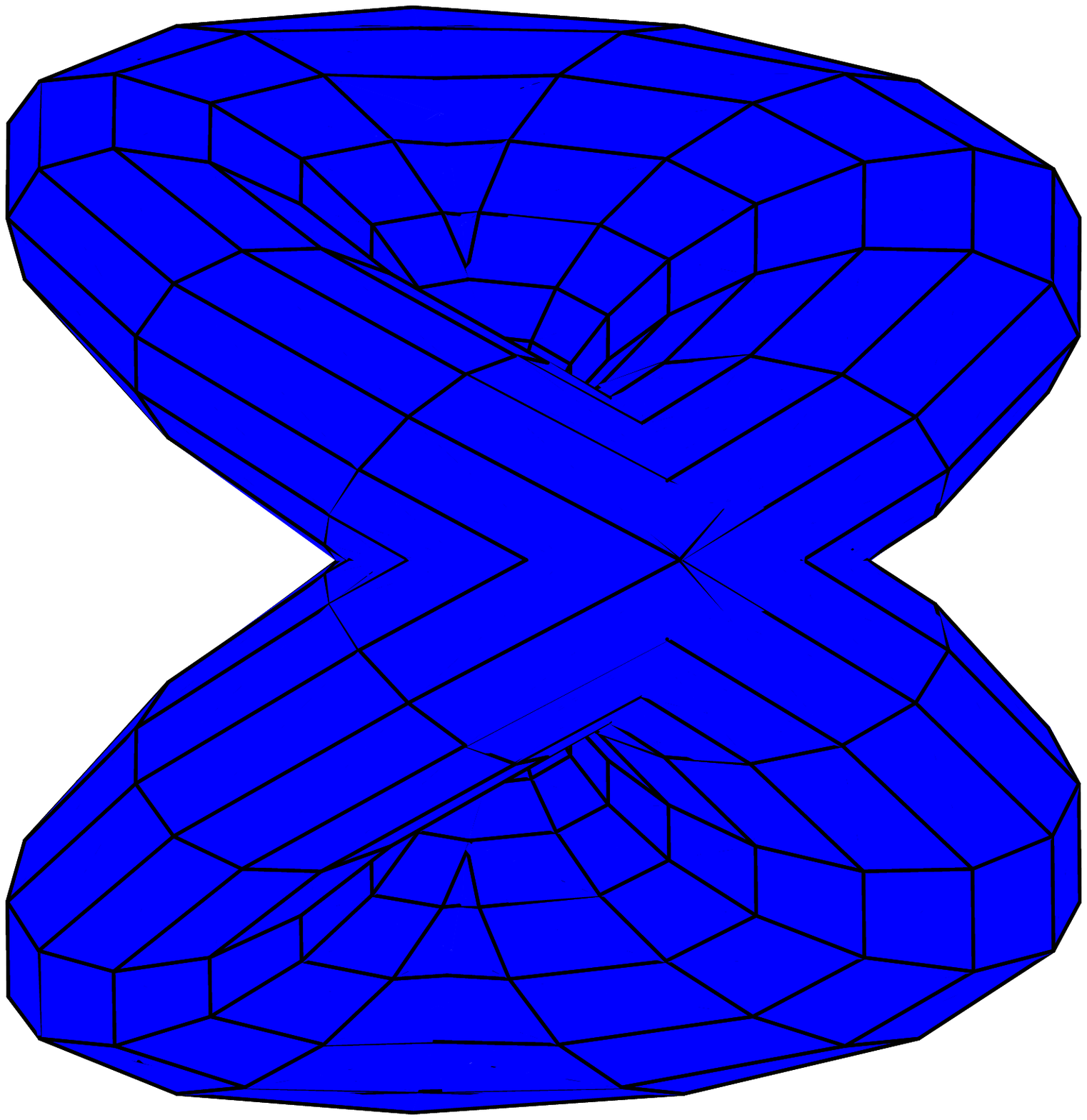}
\caption{shows the surface on which $Q_2$ lies.}
\end{figure}
\begin{figure}[htbp] \epsfxsize=4cm \epsfbox{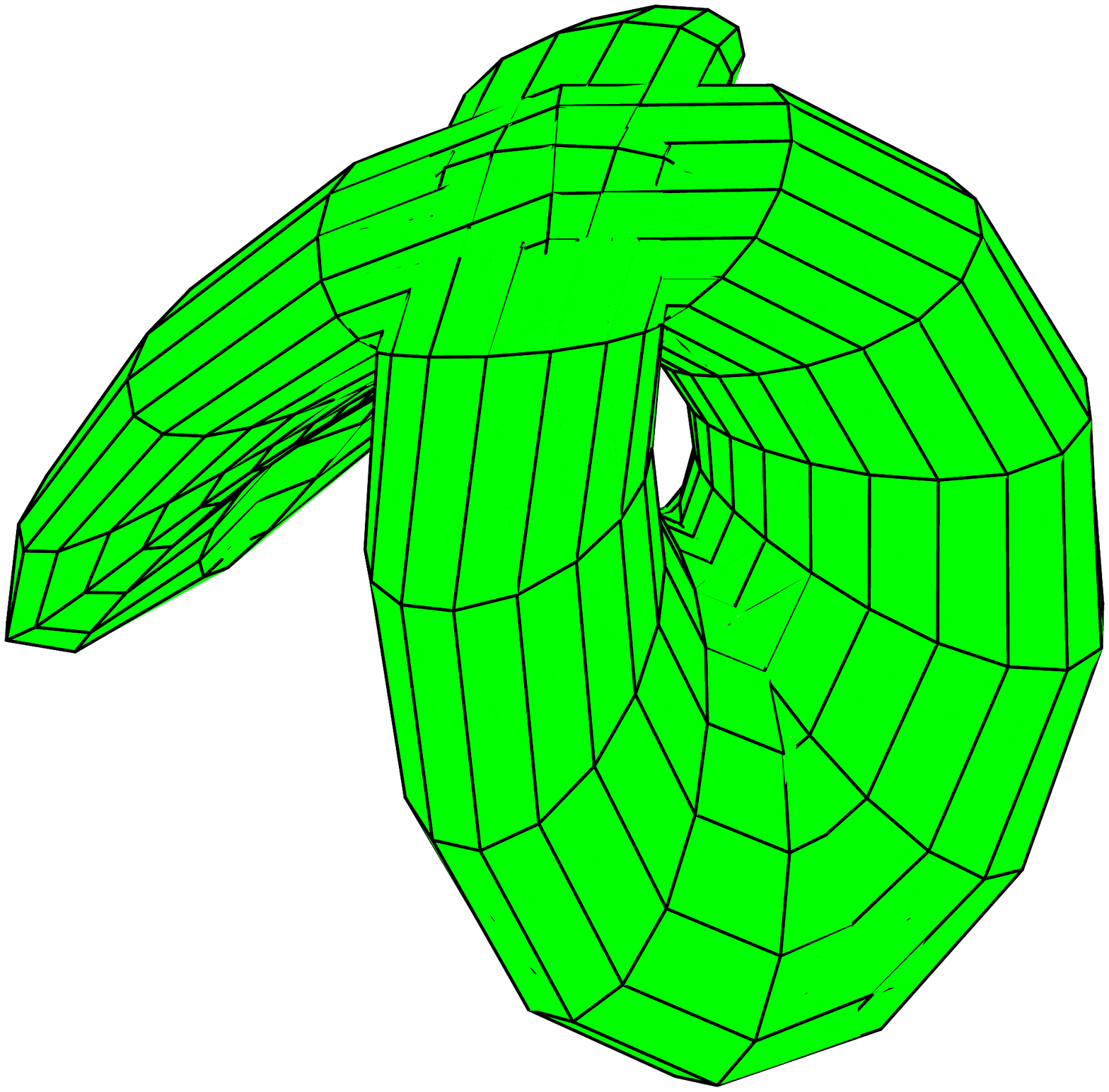}
\caption{shows the surface on which $Q_3$ lies.}
\end{figure}
\begin{figure}[htbp] \epsfxsize=4cm \epsfbox{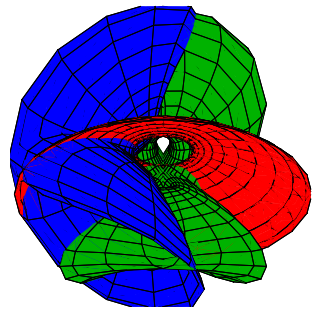}
\caption{shows the surface on which all the three vector fields, $Q_1$, $Q_2$, and $Q_3$ lie.}
\end{figure}

For each of the gauge fields ${\bf{Q}}^i$ we can find three
variables $\alpha^i$,$\beta^i$ and $\psi^i$ such that
${\bf{Q}}^i=\alpha^i{\vec{\nabla}}\beta^i+{\vec{\nabla}} \psi^i$ (no
summation in $i$). Explicit expressions are: \bea
\alpha^1&=&Tan^{-1}(\frac{z}{y})+Tan^{-1}(\frac{2x}{(1-\vec{x}^2)}) \nn \\
\alpha^2&=&Tan^{-1}(\frac{x}{z})+Tan^{-1}(\frac{2y}{(1-\vec{x}^2)}) \nn \\
\alpha^3&=&Tan^{-1}(\frac{y}{x})+Tan^{-1}(\frac{2z}{(1-\vec{x}^2)}) \nn \\
\beta_1&=&\frac{z^2+y^2}{g(1+\vec{x}^2)} ; \beta_2=\frac{x^2+z^2}{g(1+\vec{x}^2)} \nn \\
\beta_3&=&\frac{y^2+x^2}{g(1+\vec{x}^2)} \nn \\
\psi^i&=&\frac{1}{4g}Tan^{-1}(\frac{\vec{x}^2-1}{2x_i}) \eea The
representation is  clearly Clebsch like with the caveat that
$\psi^i$ is multiple valued, and therefore contributes a non
vanishing contribution to the helicity when integrated over a closed
contour. This decomposition allows us to separate the contributions
of the field and the fluid. Such structures for the pure gauge field
($A_{i}$) have been used in magnetohydrodynamics to find third order
linkages between three magnetic fields in magnetic recombination and
geophysical processes \cite{Mayer,semenov, kam}. In these works,
each component of the SU(2) field is considered to be a U(1) Abelian
magnetic field.

Once the solution with winding number $n=1$ has been found, the solutions with higher order winding numbers n can be obtained by applying  the gauge transformation  $\Omega(x)^{n-1}$ to the n=1 solution\cite{jackiwrebbi}. Thus a whole spectrum of fluid field knots with integral winding numbers can be produced. It should be empasized that these knots consist of both the velocity of the fluid and the gauge potential of the Yang-Mills field. In the region exterior to the knot, the gauge potentials have to satisfy appropriate constraints.

The solutions we have constructed, therefore, have the  intended character; they
are localized as well as topologically nontrivial. The exterior solution ($Q^{a}{}_{\mu}=\frac{m}{g}f U^{a}{}_{\mu}+ A^{a}{}_{\mu}=0$, implying $j^{a}{}_{\mu}\propto
 A^{a}{}_{\mu}$) is  the non-Abelian analog of the London equation, and  displays, what might be viewed as  an ``inverse'' Meissner effect;  the "magnetic" flux is pushed out
of the exterior region into the interior region of the knots, which can be regarded as chromomagnetic knotted flux tubes.

Because the unified connection {\bf{Q}} combines the fluid and the Yang Mills fields, the vanishing of its time component is not an empty condition; it implies that the time component of the gauge potential is proportional to the time component of the species velocity. This again can be viewed as a generalized Coulomb gauge condition providing for the staticity of the solution.
Since the Pontrayagin number labels these topological solutions, it is suggestive to consider it as a quantum number for such solutions.

There has been recent speculation that glueballs in a Yang-Mills
theory may just be  such topological, knotted solutions with their
energies providing an analog of the energy levels of Bohr's atom
\cite{Buniy,niemi,fadeev}. A detailed  estimate of the energies of
our topological solutions is yet to be carried out, but  there does
exist  a radial length scale in the theory given by $\frac{m}{g}$,
which should lead to a non zero minimum energy of these knots.

Knotted solutions, developed in this work, should exist not only in
the QGP, but also in quark stars and the early universe. A detailed
investigations of the  physical properties of these fluid field
knots figures to be important for all strongly coupled quark- gluon
matter.

\end{document}